\documentclass[journal]{IEEEtran}

\usepackage{graphicx} % Required for inserting images
\usepackage{booktabs}
\usepackage{subfig}
\usepackage{tabularx}
\usepackage{multirow}
\usepackage{amssymb} % for \checkmark
\usepackage{pifont}  % for \xmark
\usepackage{enumitem}
\usepackage{url}
\usepackage{hyperref}
\usepackage{xcolor}     % in preamble
\usepackage{makecell} % put this in your preamble
\usepackage{array} % ensures p/m/b column types work well
\newcolumntype{L}{>{\raggedright\arraybackslash}X}  % left-aligned wrapped
\newcolumntype{C}{>{\centering\arraybackslash}X}    % centered wrapped

\newcommand{\xmark}{\ding{55}} % define xmark using pifont

\newcolumntype{Y}{>{\centering\arraybackslash}X}

\usepackage{xcolor}
\usepackage[numbers,sort&compress]{natbib}

\usepackage{tikz}
\usetikzlibrary{positioning, arrows.meta, shapes.geometric, fit, backgrounds}

\usepackage{amsmath} % For mathematical symbols

\newcommand{\rev}[1]{\textcolor{black}{#1}}

\newcommand{\frameworkName}{CHIPSIM}

\title{\frameworkName{}: A Co-Simulation Framework for Deep Learning on Chiplet-Based Systems}

\author{
    \IEEEauthorblockN{Lukas Pfromm\IEEEauthorrefmark{1},
                      Alish Kanani\IEEEauthorrefmark{1},
                      Harsh Sharma\IEEEauthorrefmark{2},
                      Janardhan Rao Doppa\IEEEauthorrefmark{2},
                      Partha Pratim Pande\IEEEauthorrefmark{2},
                      Umit Y. Ogras\IEEEauthorrefmark{1}}
    \\

    \IEEEauthorblockA{\IEEEauthorrefmark{1}University of Wisconsin–Madison, Madison, WI, USA},  %\\
    \IEEEauthorblockA{\IEEEauthorrefmark{2}Washington State University, Pullman, WA, USA}
    \vspace{-5mm}
}

\begin{document}
\maketitle

\begin{abstract}
    \label{sec:abstract}
Due to reduced manufacturing yields, traditional monolithic chips cannot keep up with the compute, memory, and communication demands of data-intensive applications, such as rapidly growing deep neural network (DNN) models. Chiplet-based architectures offer a cost-effective and scalable solution by integrating smaller chiplets via a network-on-interposer (NoI). Fast and accurate simulation approaches are critical to unlocking this potential, but existing methods lack the required accuracy, speed, and flexibility.
To address this need, this work presents \frameworkName{}, a comprehensive co-simulation framework designed for parallel DNN execution on chiplet-based systems. \frameworkName{} concurrently models computation and communication, accurately capturing network contention and pipelining effects that conventional simulators overlook.
Furthermore, it profiles the chiplet and NoI power consumptions at microsecond granularity for precise transient thermal analysis. Extensive evaluations with homogeneous/heterogeneous chiplets and different NoI architectures demonstrate the framework’s versatility, up to 340\% accuracy improvement, and power/thermal analysis capability.
\end{abstract}


\vspace{-3mm}
\section{Introduction} \label{sec:introduction}

Deep neural networks (DNNs) require increasingly powerful systems due to their growing size and computational demands~\cite{sun2019summarizing}. Traditional monolithic 2D chips have grown substantially to keep up with the exponential increase in model parameters and processing requirements. However, monolithic chips becomes less economical to manufacture due to declining yield as chip size increases~\cite{cunningham1990yieldModel}. These trends have driven a new advanced packaging and heterogeneous integration approach using smaller dies called chiplets~\cite{graening2024cost, amd2021chipletTechnology}.

\begin{figure}[t]
    \centering
    \includegraphics[width=0.7\linewidth]{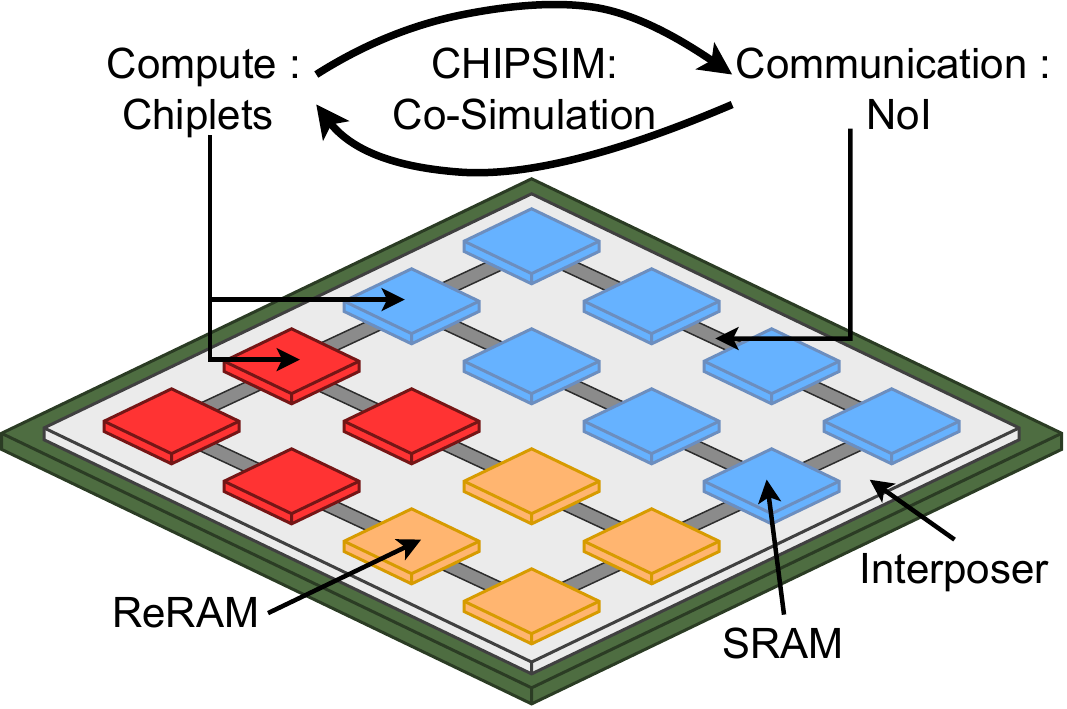}
    \vspace{-1mm}
    \caption{An illustration of the \frameworkName{} framework on a non-mesh heterogeneous chiplet-based system.}
    \vspace{-7mm}
    \label{fig:chiplet_system}
\end{figure}

Chiplet-based systems integrate many chiplets into a single package, as illustrated in Figure~\ref{fig:chiplet_system}. 
Each chiplet has a lower manufacturing cost than a monolithic chip due to its small size and higher yield \cite{stow2016costAnalysis, graening2024cost}. Combining chiplets into a single package can deliver higher performance and energy efficiency than monolithic systems at comparable costs. 
Moreover, the greater modularity and heterogeneity enrich the design space, enabling systems tailored to specific application requirements \cite{sharma2025hemu}.

Chiplet-based systems have experienced widespread adoption in both academia and industry. Studies in this area have explored various topics, including network topology, packaging, interconnects, power and thermal analysis, and application mapping \cite{kanani2025thermos}. In parallel, industry adoption has also accelerated, with notable examples such as NVIDIA’s B200 \cite{goldwasser_gb200_2024}, Intel’s Agilex and Meteor Lake \cite{intel2024agilex, gomes2022meteor}, AMD's Ryzen, EPYC, and Instinct \cite{amd2021chipletTechnology}, and other commercial products built on chiplet architectures. Federal funding initiatives, like the National Advanced Packaging Manufacturing Program (NAPMP) and Scalable Memory Architecture Program, have also been launched to support chiplet-based system development \cite{nist_napmp, nist_smap}. This convergence of academic research, industrial implementation, and strategic initiatives underscores the importance of chiplets and the supporting tools for their development. 
%As adoption grows, there is a pressing need for effective evaluation methods.
 
Chiplet-based systems require fast and accurate evaluation methods to quantify the trade-off between competing objectives, such as performance, energy efficiency, and temperature. Simulations are essential throughout the design process, from early exploration and chiplet selection to floorplanning and workload mapping, especially when hardware prototypes are unavailable. 
Navigating complex and conflicting constraints to meet the design objectives is intractable without reliable tools. 
Large-scale heterogeneous architectures and target workloads further increase the complexity. As design complexity grows, fast and accurate simulation is key to enabling iteration, guiding decisions, and driving innovation. 

\begin{figure}[t]
    \centering
    \includegraphics[width=0.9\linewidth]{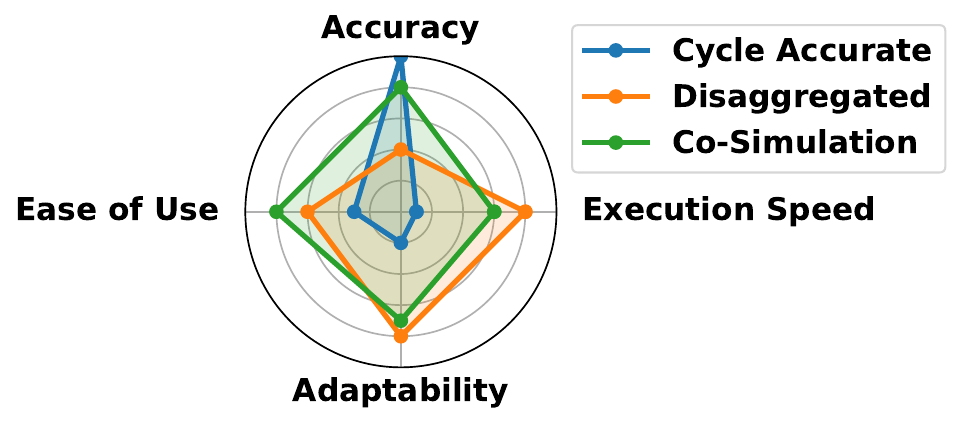}
    \vspace{-2mm}
    \caption{A visual comparison between existing simulators and the proposed simulation methodology.}
    \vspace{-6mm}
    \label{fig:simulator_comparison}
\end{figure}

Current research on chiplet-based systems relies on simulation methods that are accurate but slow and inflexible or fast but inaccurate, as illustrated in Figure~\ref{fig:simulator_comparison}. For instance, when cycle-level accuracy is needed, gem5~\cite{lowe2020gem5,binkert2011gem5} can simulate the computations, while HeteroGarnet~\cite{bharadwaj2020kite} or Booksim~\cite{jiang2013booksim} simulates communication. However, gem5 is slow and has limited support for alternative processing technologies beyond CMOS designs. 
To achieve faster simulations or explore novel configurations such as chiplets based on in-memory computing (IMC), researchers frequently combine existing simulators in an ad-hoc fashion. Specifically, \textit{the computation and communication phases of the target applications are considered separately, with limited interaction between them}~\cite{iff2023hexamesh, sharma2023florets, sharma2022swap, iff2023rapidchiplet, li2022gia}. 
Improvised methods cannot accurately capture complex behaviors like traffic contention and pipelining of DNN model layers, which can impact the simulation performance results by more than 340\%, as we demonstrate in Section~\ref{sec:results}.
Moreover, decoupling computation and communication prevents \textbf{a coherent notion of time}, making modeling time-based behavior such as transient power analysis impossible.
Thermal analysis, a crucial aspect of chiplet-based systems, requires time-based power consumption. Therefore, chiplet-based systems need fast, accurate, and modular co-simulation approaches.

\rev{This paper introduces \frameworkName{}, a co-simulation framework for chiplet-based systems executing DNN models. While co-simulation has been explored in other domains, \frameworkName{} is the first framework to apply this methodology to chiplet-based architectures. In addition to co-simulation, \frameworkName{} coordinates computation and communication phases under a unified notion of global simulation time, enabling concurrent execution of multiple DNN models while accurately capturing their interactions and system-level effects.}
\frameworkName{} begins by simulating the first layer of the target DNN workload. Upon completion, it simulates the communication of the resulting activations to the chiplet that will run the consecutive layer. This way, computation and communication events are simulated in tandem while keeping track of the global time.
\rev{When multiple DNN models arrive in a stream and must execute in parallel on the same system, \frameworkName{} orchestrates concurrent computation and communication simulations while accurately capturing their interactions and impacts on each other.} For example, the communication simulation at any given time accounts for all active chiplet-to-chiplet data movement of all DNNs in the system to capture network contention accurately. 
\textit{Our results show contention becomes a dominant factor in communication latency, especially during highly utilized times, yet it remains unmodeled by current approaches}. 

As a critical and novel aspect, \frameworkName{} incorporates a detailed power tracking mechanism that generates power and thermal profiles with microsecond-level granularity for every chiplet. These power profiles enable transient and steady-state thermal analysis. 
Additionally, \frameworkName{} supports both heterogeneous chiplet types and custom network-on-interposer (NoI) topologies to support extensive design-space exploration and diverse configurations. 
Due to its modularity, \frameworkName{} can utilize different computation and communication simulators. \rev{In this work, we primarily target IMC chiplets simulated using CimLoop~\cite{andrulis2024cimloop} and use HeteroGarnet~\cite{bharadwaj2020kite} for communication. However, we perform additional hardware validations where CiMLoop is replaced with an analytical compute model. We emphasize that \textit{the proposed framework is not limited to any processing technology or choice of simulators}.}

The key contributions of this work are as follows:
\begin{itemize}[leftmargin=*]
    \item A comprehensive co-simulation framework that captures the salient features of hardware execution by modeling computation and communication events,
    \item A concrete realization with open-source tools and modular support for heterogeneous chiplets and NoI topologies, 
    \item Comprehensive evaluations with three different system configurations that demonstrate up to 340\% higher accuracy and enable transient power/thermal analysis—capabilities not possible with existing approaches.
    \item Open-sourced code for the created tool available at \href{https://github.com/LukasPfromm/CHIPSIM}{github.com/LukasPfromm/CHIPSIM}.
\end{itemize}

The rest of the paper is organized as follows. Section \ref{sec:related_work} reviews the related work. Section \ref{sec:simulation_framework} details our approach to co-simulation, while Section \ref{sec:concrete_implementation} presents a concrete realization. Finally, Section \ref{sec:results} presents the experimental evaluation, and Section~\ref{sec:conclusion} concludes the paper.
\vspace{-3mm}
\section{Related Work} \label{sec:related_work}

\begin{table*}[t]
\centering
\caption{Feature comparison of simulation frameworks.}
\scriptsize
\resizebox{\textwidth}{!}{%
\begin{tabular}{l|ccccccc}
\toprule
\textbf{Simulator / Analysis Tool} & 
\textbf{Sim. Time} & 
\begin{tabular}[c]{@{}c@{}}\textbf{Thermal}\\\textbf{Analysis}\end{tabular} & 
\begin{tabular}[c]{@{}c@{}}\textbf{Parallel Model}\\\textbf{Execution}\end{tabular} & 
\begin{tabular}[c]{@{}c@{}}\textbf{Pipeline}\\\textbf{Support}\end{tabular} & 
\begin{tabular}[c]{@{}c@{}}\textbf{Network}\\\textbf{Contention}\end{tabular} & 
\begin{tabular}[c]{@{}c@{}}\textbf{Heterogeneous}\\\textbf{Systems}\end{tabular} & 
\begin{tabular}[c]{@{}c@{}}\textbf{IMC}\\\textbf{Support}\end{tabular} \\
\midrule
Gem5 \cite{lowe2020gem5, binkert2011gem5} + HeteroGarnet \cite{bharadwaj2020kite} & Weeks & \xmark & \checkmark & \checkmark & \checkmark & \checkmark & \xmark \\
SIAM \cite{krishnan2021siam} & Hours & \xmark & \xmark & \xmark & \checkmark & \xmark & \checkmark \\
HISIM \cite{wang2025hisim} & Seconds & \checkmark$^*$ & \xmark & \xmark & \xmark & \checkmark & \checkmark \\
\textbf{\frameworkName{} (This Work)} & Minutes & \checkmark & \checkmark & \checkmark & \checkmark & \checkmark & \checkmark \\
\bottomrule
\end{tabular}%
}
\vspace{-5mm}
\label{tab:sim_feature_comparison}
\end{table*}

Chiplet-based system design presents many challenges, from optimizing individual chiplets to NoI design and mapping workloads onto these systems. Many studies have leveraged the modularity enabled by chiplet integration to develop both general-purpose and heterogeneous systems tailored to specific domains, such as AI/ML~\cite{amd2021chipletTechnology, krishnan2022big, sharma2025heterogeneous, smith2024realizing_mi300, jaiswal2024halo, gomes2022meteor}. Similarly, the design of the chiplet interconnection network has attracted significant attention due to its importance in minimizing latency and energy cost incurred by data movement \cite{bharadwaj2020kite, iff2023hexamesh, sharma2023florets, sharma2022swap}. To maximize the performance of these systems after their design, recent studies have proposed methods for mapping computation and communication tasks \cite{shao2019simba, sharma2023florets} and optimizations such as DNN layer pipelining \cite{song2017pipelayer, shao2019simba, odema2024scar}. 

Solving these design challenges requires fast, flexible, and accurate simulation models to capture the complex interactions inherent to these architectures. They must adapt to diverse aspects of system design, from chiplet composition and interconnection topology to workload mapping and runtime optimizations, while executing quickly enough to enable broad design space exploration. However, existing simulation approaches fall short in one or more of these aspects, limiting their effectiveness in supporting comprehensive analysis.

Studies that focus purely on NoI design sometimes omit computation performance modeling~\cite{sharma2023florets, li2022gia, iff2023hexamesh, kannan2015enabling, iff2023rapidchiplet} or perform decoupled compute and communication simulation~\cite{sharma2022swap, krishnan2021siam, sharma2025heterogeneous}. While simplifying modeling, this approach introduces inaccuracies since compute-communication interactions are abstracted. 
Table~\ref{tab:sim_feature_comparison} compares the features supported by several proposed simulators. Gem5 \cite{lowe2020gem5, binkert2011gem5}, while not created specifically for chiplets, is a cycle-accurate simulator that has been adapted via the HeteroGarnet simulator \cite{bharadwaj2020kite} to model inter-chiplet traffic. While accurate, cycle-accurate simulation comes at the cost of prohibitive runtime. Additionally, there is no native support for thermal modeling or processing approaches such as IMC \cite{andrulis2023raella, wan2022compute}. 
SIAM~\cite{krishnan2021siam} aims to address the need for IMC chiplet simulation by using NeuroSim \cite{peng2019neurosim} for processing simulation in combination with Booksim \cite{jiang2013booksim} for traffic modeling. However, it does not support heterogeneous systems or thermal analysis and still requires hours for simulation. To mitigate the high simulation times of previous approaches, HISIM \cite{wang2025hisim} utilizes analytical models instead of existing simulators. This allows for a considerable simulation time improvement to the order of milliseconds or seconds. Additionally, thermal modeling is implemented along with native support for IMC chiplets. However, rapid analysis comes at the cost of accuracy and features. Thermal analysis is not transient; instead, discrete thermal simulations are performed for each model layer.

Neither SIAM nor HISIM support the modeling of (1) pipelined behavior, where two or more layers of a model execute concurrently, (2) parallel model execution, where two or more models run simultaneously on the same system, or (3) communication contention, which occurs on the NoI. \rev{Therefore, they cannot accurately model the scenario where a stream of models is executed in parallel in the same system, a workload commonly applied to chiplet-based systems \cite{sharma2023florets,amd2021chipletTechnology,shao2019simba}, while cycle-accurate simulations are too slow.} \rev{To meet this need, we apply co-simulation to the domain of chiplet-based system simulation and create \frameworkName{}. We note that co-simulation is an established methodology across multiple domains, including end-to-end networked systems (e.g., SimBricks \cite{li2022simbricks}) and processor–programming model studies (e.g., SST \cite{rodrigues2011structural}). \frameworkName{} builds on this foundation by adapting time-synchronized simulation to the domain of chiplet-based systems executing parallel DNN workloads. Unlike prior frameworks, \frameworkName{} integrates compute and communication simulators tailored to chiplet architectures, supports multi-instance DNN pipelines, and provides microsecond-resolution thermal analysis by tracking energy and execution timelines across chiplets. This domain-specific instantiation enables accurate performance, power, and thermal modeling in emerging heterogeneous chiplet+NoI designs that are not well supported by existing general-purpose simulation infrastructures.} 
Our evaluations in Section~\ref{sec:results} compare \frameworkName{} against two baseline approaches: communication simulation only and decoupled computation-communication modeling (like HISIM and SIAM). The results highlight the need for a fast, accurate, and modular co-simulation framework to integrate computation and communication models coherently.
\frameworkName{} addresses this need as a comprehensive framework for detailed performance, power, and thermal analysis.

\vspace{-3mm}
\section{\frameworkName{} Chiplet Co-simulation Framework} \label{sec:simulation_framework}

\begin{figure*}[t]
    \centering
    \includegraphics[width=\textwidth]{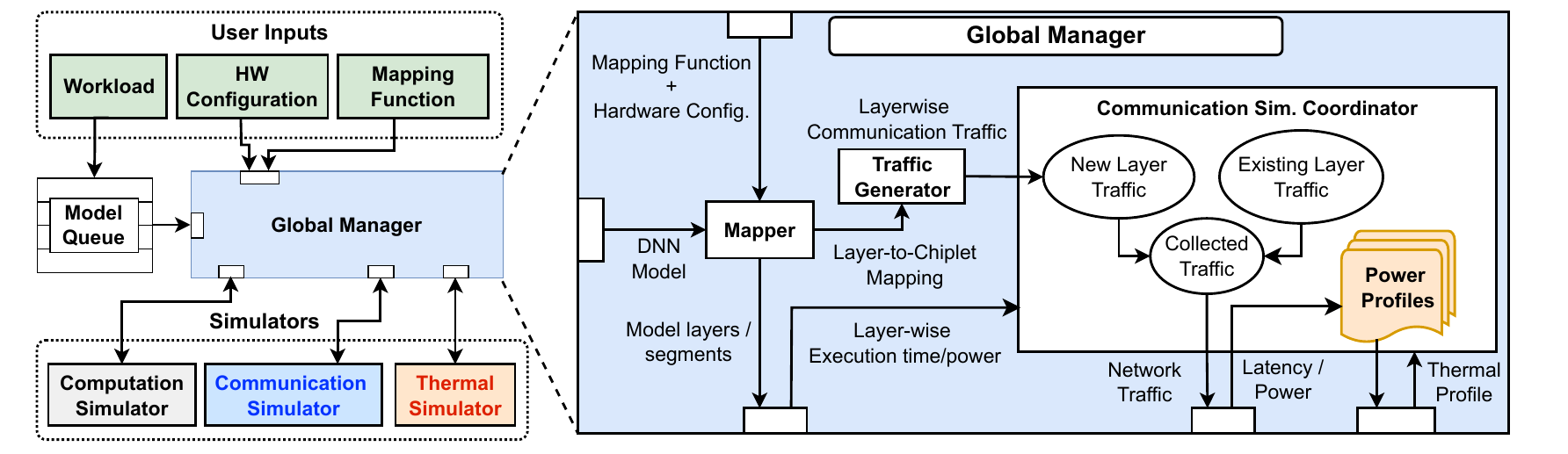}
    \vspace{-5mm}
    \caption{Outline of the proposed simulation framework.}
    \vspace{-5mm}
    \label{fig:system_diagram}
\end{figure*}

\subsection{Overview of the Proposed Framework} \label{ssec:simulation_framework}

DNN models follow a layer-wise execution: (1) the computations of the current layer are executed, (2) the generated activations are transmitted to the hardware resources that will execute the subsequent layer. These computation--communication steps are repeated until the entire model is complete.
The \frameworkName{} framework accurately models these interactions between computation and communication while multiple DNN model instances run in parallel by tracking the compute and communication phases of each model individually. 
Additionally, \frameworkName{} tracks the occupancy of each chiplet, since the mapping of subsequent DNN model layers depends upon the execution of previous layers. 
Tracking detailed utilization statistics can also aid workload mapping and architecture optimization techniques. 

The building blocks of the \frameworkName{} framework and data flow between them are visualized on the left-hand side of Figure~\ref{fig:system_diagram}. 
There are three primary user inputs: 
(1) Target DNN workload, 
(2) Hardware configuration, 
(3) Mapping function of the input DNN models to the hardware resources.

The core component of \frameworkName{}, the \texttt{Global Manager}, takes these inputs and orchestrates the computation and communication co-simulation as shown in Figure~\ref{fig:system_diagram}. 
The workload comprises a stream of DNN models, which is continuously read and mapped to the target chiplet-based system by the \texttt{Global Manager}.
The mapping strategy for DNN layers is determined by the \texttt{Mapping Function} input.
The hardware configuration describes the number and type of chiplets, compute capabilities, memory capacities, and the NoI topology. 
After mapping, the \texttt{Global Manager} invokes the computation and communication simulators to collect power and performance metrics for each DNN model as a function of time. 
This section provides a conceptual description of the proposed \frameworkName{} framework. 
This concept can be realized with available compute \cite{andrulis2024cimloop, samajdar2018scale, lowe2020gem5} and communication \cite{jiang2013booksim, bharadwaj2020kite, feng2024cnsim, catania2015noxim} simulators, as described in Section~\ref{sec:concrete_implementation}.

\vspace{-3mm}
\subsection{Input Workload and Mapping} \label{ssec:input_and_mapping}
\vspace{-1mm}

The input workload is a list of DNN models that will run on the target chiplet-based system. The DNN models are represented in a layer-wise format, where each layer is characterized by its parameters, including the number of input and output layers, filter sizes, and stride lengths.

The \texttt{Global Manager} traverses the \texttt{Model Queue} to find the first DNN model that fits the available memory resources in the target hardware. We allow out-of-order execution of the DNN models to maximize chiplet utilization and prevent smaller models from starving when they arrive after a large model. We emphasize that the \frameworkName{} co-simulation framework is oblivious to the specific arbitration method.

After retrieving the next DNN model, the \texttt{Global Manager} utilizes the user-provided \texttt{Mapping Function} to allocate \textit{each layer of this DNN model} to the available resources in the chiplet-based system. 
Like the arbitration mechanism, the \frameworkName{} framework can be used with any mapping function. 
To maximize generality, \frameworkName{} supports layers that may be too big to fit on one chiplet. In this case, the layer is split into multiple \textit{segments}, with each segment being mapped to an individual chiplet.
As the \texttt{Global Manager} traverses each layer of the current DNN sequentially, it maps the next layer to a chiplet if there is enough available memory. Otherwise, it divides the layer into the fewest segments that fit the chiplet resources and maps them to minimize the communication cost.
Then, it updates the system state to keep track of the memory resource usage in each chiplet. This update ensures that the occupancy of each chiplet is accurate for future DNN mapping operations. 

\vspace{-3mm}
% \subsection{Compute Simulation}

% Compute operations within each chiplet are independent of the compute operations of every other chiplet. Therefore, the \texttt{Global Manager} launches a new computation simulation thread for each new layer mapped into one or multiple chiplets (if segmented) to accelerate the simulation process. 
% Hence, the number of compute simulator instances is equal to the number of model segments. The \texttt{Global Manager} can easily be interfaced with any computation simulator, as demonstrated in Section~\ref{sec:concrete_implementation}.

% Each simulation thread takes the physical system configuration, which defines the properties of each chiplet, along with the model segment. 
% Then, it returns the total execution time, energy, and power consumption for the compute operation. 

\subsection{Compute Simulation}
Compute operations within each chiplet are independent of the operations executed on other chiplets. To exploit this parallelism, the \texttt{Global Manager} launches a dedicated compute simulation thread for each layer segment mapped to one or more chiplets. Hence, the number of active compute simulator instances equals the number of model segments. Each thread is provided with the physical system configuration—defining chiplet properties such as MAC units, memory hierarchy, and frequency, alongside the DNN layer description (e.g., type, input/output dimensions). The simulator then returns the estimated execution time, energy, and power consumption.

\rev{To ensure extensibility and long-term usability, \frameworkName{} is designed with a modular interface that decouples the coordinator from specific backends. Each component simulator is invoked independently through a standardized input/output format. For compute simulation, this abstraction allows seamless integration with a range of backends, including analytical models, cycle-accurate simulators, or third-party tools. For example, we replaced the original CiMLoop-based backend with an analytical CPU performance model that estimates compute latency by dividing the number of MAC operations by the sustained throughput (MACs per second) of the target CPU, as shown in Section \ref{ssec:hw_validation}. Importantly, this substitution requires no modifications to the control flow or data structures of \frameworkName{}, demonstrating the generality of the framework. Since simulation progress is decoupled from the internal execution model of each backend, \frameworkName{} preserves fidelity while supporting both rapid prototyping and detailed architectural studies.}

\vspace{-3mm}
\subsection{Communication Simulation}
When a compute simulation thread completes, it produces layer-wise compute metrics for that layer as depicted in Figure \ref{fig:system_diagram}. Layer-wise activations have been previously generated by the traffic generator.  
The communication simulation is responsible for accurately tracking the communication latency and energy cost. 
Unlike the computations within each chiplet, \textit{the inter-chiplet communication network is a shared resource used by multiple DNN models simultaneously.} 
Hence, there is a single communication simulation thread that accounts for all inter-chiplet communication from all active DNN models. 
The \texttt{Global Manager} generates the communication traffic for all models and inputs them into a \texttt{Communication Simulation Coordinator}, as shown in Figure~\ref{fig:system_diagram}.

Consider a toy example where the workload consists of only a single DNN model. 
The \texttt{Communication Simulation Coordinator} would simply dispatch each layer's activation traffic to the network simulator layer by layer. However, since multiple DNN models may run in parallel in a server-class processor, the \frameworkName{} framework also handles much more complex cases with tens of concurrent DNN models. To do this, the \texttt{Communication Simulation Coordinator} tracks \textbf{\textit{all other active DNN models and their traffic at a given time}}.
Using the results of the compute simulation operations, the traffic between each chiplet, and the current state of the simulated system, the \texttt{Communication Simulation Coordinator} dispatches communication operations to the network simulator. This coordination allows for the modeling of the contention between layer traffic, as detailed in the following section.

\vspace{-3mm}
\subsection{Co-simulation Coordination} \label{ssec:simulation_coordination}

We employ an illustrative scenario to describe how \frameworkName{} accurately models the behavior of a chiplet-based system.

\begin{figure} [t]
    \centering
    \includegraphics[width=1\linewidth]{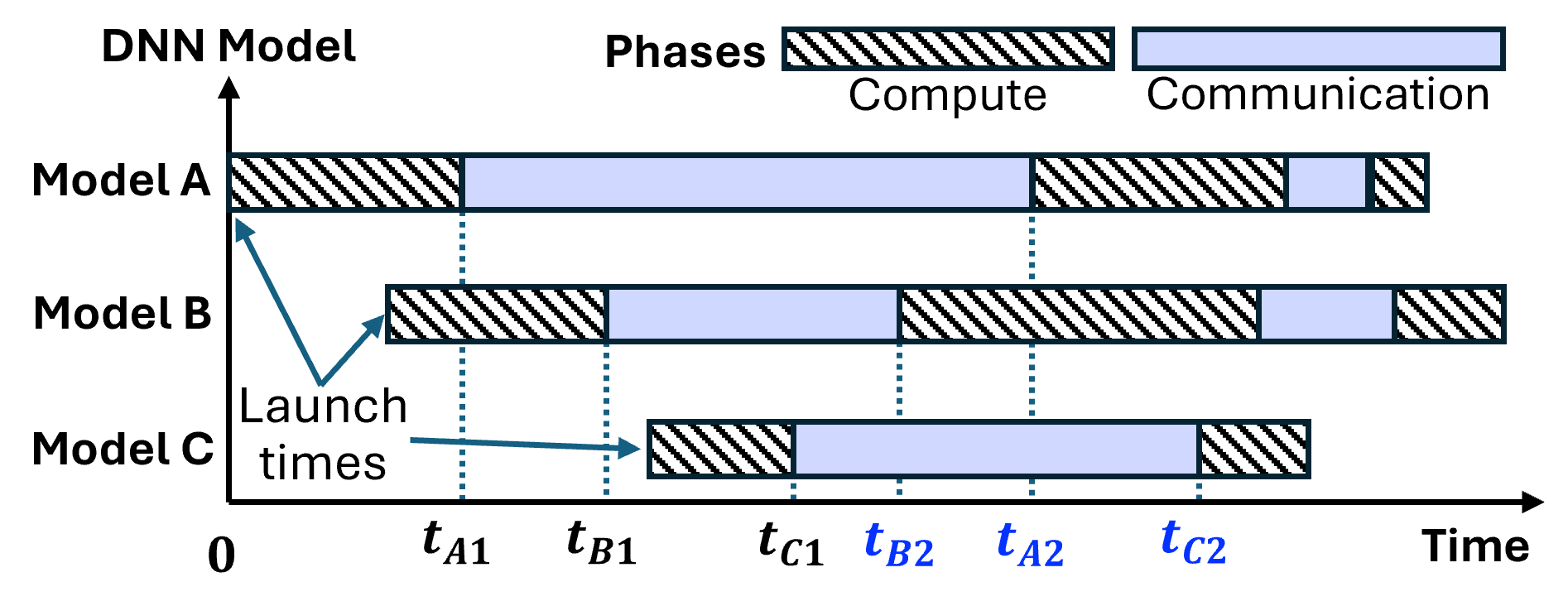}
    \caption{An illustrative event diagram showing the interleaving of different simulation operations.}
    \vspace{-6mm}
    \label{fig:event_diagram}
\end{figure}

\noindent\textbf{Scenario description:}
Three DNN models run in parallel, as shown in a different row along the y-axis in Figure~\ref{fig:event_diagram}. The x-axis shows the simulation time and labels time instances that are important for this discussion.
Model A is launched at time 0, while Model B and Model C are launched later with arbitrary offsets, as marked in the figure. 
For simplicity, Models A and B have three layers, while Model C has two layers.
Each model layer alternates between compute and communication phases, corresponding with computing the activations of a given layer and communicating them to the chiplets running the next layer.
These phases are shown by black diagonal and light blue plain patterns in Figure~\ref{fig:event_diagram}. 
Next, we describe the co-simulation behavior step by step using the event diagram.

\vspace{2pt}
\noindent\textbf{Time $0- t_{A1}$:} 
At time 0, Model A simulation begins, and after an offset, Model B also starts running. Since no communication simulations are performed until $t_{A1}$, there is no data movement (hence no communication simulation) in this interval.

\vspace{2pt}
\noindent\textbf{Time $t_{A1}-t_{B1}$:} 
After Layer-1 of Model A finishes computation at time $t_{A1}$, the \texttt{Global Manager} receives the latency and energy for the operation. Additionally, the number of activations between Layer-1 and Layer-2 is known from the \texttt{Traffic Generator}.
The chiplets that execute the consecutive layer are known through the \texttt{Mapping Function}. 
So, the \texttt{Global Manager} generates the network traffic from the chiplet(s) that run Layer-1 (source) to the chiplet(s) that run Layer-2 (destination). 
Since there is no other active communication during this time, the communication simulation is launched with this traffic. 
The \texttt{Global Manager} uses the end-to-end communication latency to schedule the computation simulation of Layer-2 of Model A,  after the projected communication time passes. 
% \textit{However, the computations of the first layer of Model B finish at time $t_{B1}$, i.e., before Model A Layer 1--Layer 2 communication.}
% Therefore, the proposed \frameworkName{} framework accounts this overlap as described next.
Since the first layer of Model B finishes computation at $t_{B1}$, i.e., before Model A Layer 1--Layer 2 communication, the communication simulation is updated to account for this overlap, as described next.

\vspace{2pt}
\noindent\textbf{Time $t_{B1}-t_{C1}$:}
The computation of Layer-1 of Model B finishes at $t_{B1}$. As with Model A in the previous time interval, the \texttt{Global Manager} receives the activations produced by the first layer. 
Then, it generates the network traffic from the chiplet(s) that run Layer-1 to the chiplet(s) that run Layer-2.
In contrast to the previous interval, two communication simulations overlap: (1) the remaining portion of Model A Layer 1--2 and (2) Model B Layer 1--2. 

Accounting for this overlap is critical to accurately modeling congestion in the inter-chiplet network. 
Therefore, \texttt{Global Manager} maintains a traffic profile that combines the existing traffic and the new layer traffic created by Layer-1 of Model B. Specifically, it updates the communication simulation to include the remaining part of the existing traffic and the newly generated data. The communication simulation resumes after the combined traffic profile has been created.  
This communication traffic update process repeats when a compute layer finishes and new activations are present. 
Similarly, when a certain communication event finishes, this process is paused to get the updated end-to-end communication latency.

\vspace{2pt}
\noindent\textbf{Time $t_{C1}-t_{B2}$:}
The first compute phase of Model C completes at $t_{C1}$. The exact process described for the previous interval is repeated. 
That is, the \texttt{Global Manager} combines 
the activations produced by Model C with existing data flows to generate the updated network traffic. As with Model B, the \texttt{Global Manager} accounts for existing traffic generated by Model A and B. 
% At $t_{C1}$, the first communication phase of Model A is about 50\% complete, while the first communication phase of Model B is about 60\% complete. Just as at the previous time interval, the traffic of each of these existing models is scaled according to it's completion percentage and combined with the new traffic generated by Layer-1 of Model C to create a traffic profile of all active networks at $t_{C1}$. 
Then, the communication simulation resumes with this combined traffic profile. 
%The result of the simulation is the end-to-end communication latency of Layer 1, Model C, which established the time point $t_{C2}$.

\vspace{2pt}
\noindent\textbf{Time $t_{B2}-t_{A2}$ and later:}
$t_{B2}$ is the time instance when the Model B Layer 1--Layer 2 communication finishes. So, the \texttt{Global Manager} launches Model B Layer 2 computations at this time. 
Similarly, Model A communication finishes, and its Layer 2 computation starts at $t_{A2}$. 
The \texttt{Global Manager} repeats the processes described so far continuously until all models and layers are fully executed.
\section{Concrete Methodology Implementation} \label{sec:concrete_implementation}

The \frameworkName{} framework supports flexible integration of different simulators to match the chiplet communication architectures under study.
This section demonstrates a concrete implementation using open-source tools, as shown in Figure~\ref{fig:concrete_instance}. 
%We emphasize that our contribution is the broadly applicable co-simulation methodology and its implementation by the \texttt{Global Manager}. 

% \question{\subsection{Cycle-Accurate vs Analytical Simulation}}
% \question{If the current explanation is not clear, we can add a proceeding section here which explains in more detail the importance of using cycle accurate simulation for communication and the lesser need for accuracy when doing computation simulation.}

%\vspace{-3mm}
\subsection{Computation Simulation}
Since our experiments focus on IMC chiplets, we adopt CiMLoop~\cite{andrulis2024cimloop} as the computation simulator. CiMLoop is selected since it enables modeling a wide range of IMC architectures accurately, with detailed crossbar arrays, peripheral circuits, and analog-digital interfaces, making it well-suited for IMC-based DNN accelerators~\cite{wan2022compute, andrulis2023raella}. CiMLoop is built on the existing and proven tools Timeloop \cite{parashar2019timeloop} and Accelergy \cite{wu2019accelergy}. Performance is modeled statistically rather than cycle-based for speed, while accuracy is maintained by operand-dependent simulation. The simulator itself is validated against the existing NeuroSim tool~\cite{peng2019neurosim}. 

CiMLoop is distributed as a Docker container with a custom API interface to preserve the overall framework's modularity. We employ this interface to dispatch computation tasks to CiMLoop and retrieve results efficiently without introducing Docker-specific dependencies into the core simulation environment. This design choice facilitates the seamless replacement or addition of alternative compute simulators. 
Future work can easily integrate simulators for traditional CPU/GPU-based processing or systolic arrays \cite{samajdar2018scale} using our example implementation.

% \rev{CiMLoop is not a cycle-accurate simulator and therefore sacrifices a small amount of accuracy as compared to cycle-accurate methods. In return, CiMLoop has a much lower simulation time. This is an acceptable tradeoff for computation simulation, as the computation activity of each chiplet is independent from each other chiplet. Therefore, the co-simulation approach applied in this work will not introduce additional error as compared to doing a non-co-simulation, disaggregated simulation.}

\vspace{-3mm}
\subsection{Communication Simulation}
Accurate communication simulation is a key component in modeling the NoI and the associated data movement. As shown in Section \ref{sec:results}, communication time contributes significantly to overall execution time. Therefore, inaccurate modeling of the inter-chiplet communication greatly impacts overall accuracy. Moreover, accurate time progression cannot be modeled if inter-chiplet communication overhead is neglected or a lumped communication model decoupled from the DNN model layer computation is used. Since power and thermal analysis requires time-based metrics, co-simulating data movement with computation is required.

\begin{figure} [t]
    \centering
    \includegraphics[width=1\linewidth]{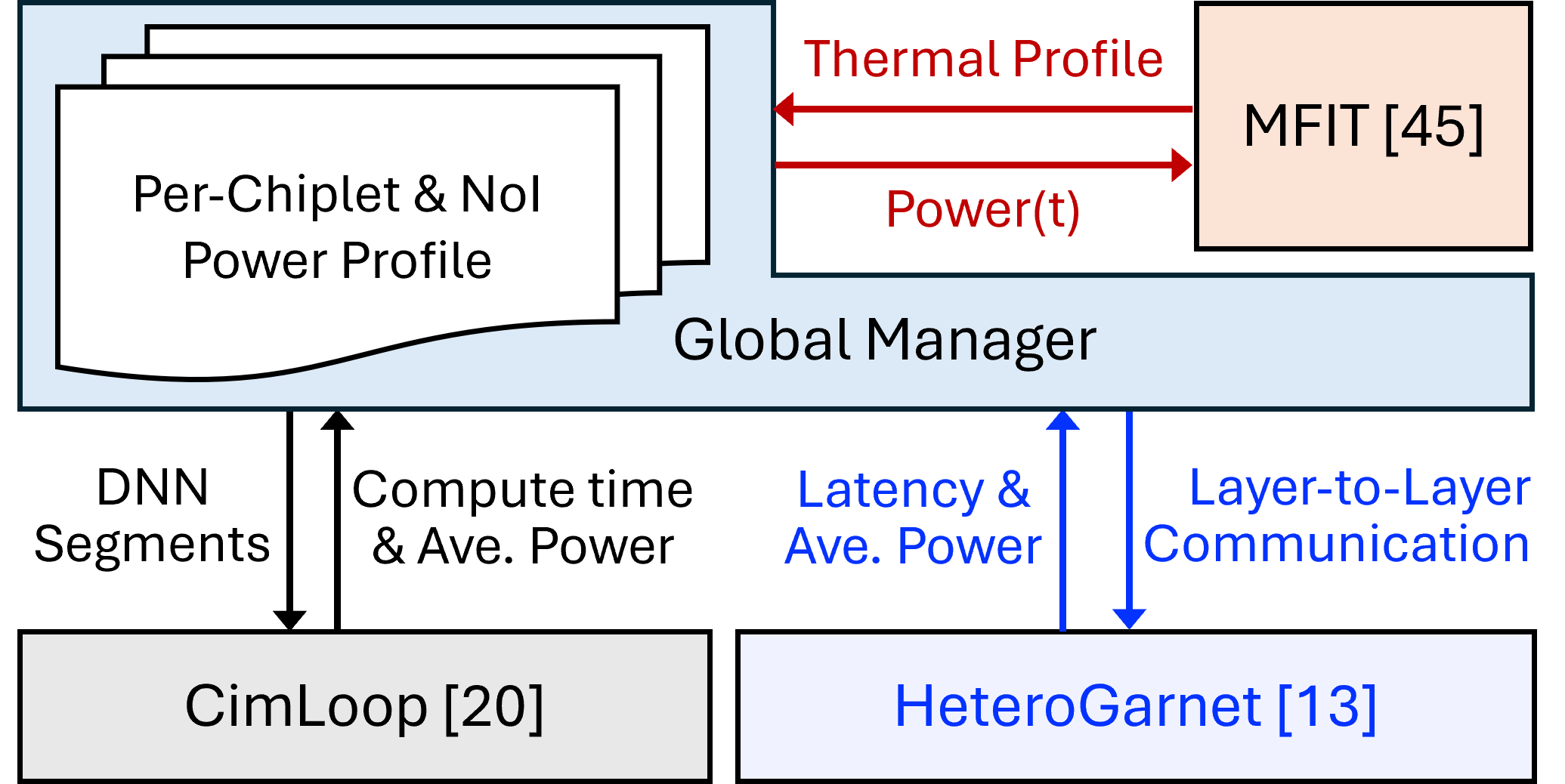}
    \caption{The illustration of our concrete implementation and interface with thermal simulations.}
    \vspace{-5mm}
    \label{fig:concrete_instance}
\end{figure}

A broad range of communication simulators have been proposed to date \cite{bharadwaj2020kite,jiang2013booksim,catania2015noxim,millberg2004nostrum,abad2012topaz}. 
The selected simulator must have certain functionality to simulate a chiplet-based system. This includes the ability to simulate large systems (hundreds of chiplets), non-homogeneous links and clock domains, custom network topologies, and support for emerging interconnect standards such as UCIe~\cite{sharma2022universal}.
We considered the alternatives listed in Table~\ref{tab:noc_sim_selection} and selected HeteroGarnet since it satisfies these requirements~\cite{bharadwaj2020kite}. 
It was developed as a modification of the existing Garnet simulator, which was created for gem5 \cite{lowe2020gem5}. \rev{Unlike the event-based CiMLoop simulator, which was selected for computation simulation, HeteroGarnet is a cycle-accurate simulator. 
As the computations of each chiplet are independent from other computations, individual event-based compute simulation does not accumulate additional modeling error. In strong contrast, layer-to-layer communications contend for shared networking resources. Hence, their detailed cycle-accurate simulation is critical for capturing the interaction of different DNN models executing in parallel.}

To integrate HeteroGarnet with the \frameworkName{} framework, we further modified it with a more user-friendly method of topology description, making it simpler to create custom topologies in the network simulator itself, along with tracking for detailed simulation metrics associated with the network traffic injected by each source. 

\begin{table}[h]
\centering
\caption{NoC simulator feature comparison.}
\scriptsize
\resizebox{\linewidth}{!}{%
\begin{tabular}{@{}l|ccccc@{}}
\toprule 
\textbf{NoC Simulator} & 
\begin{tabular}[c]{@{}c@{}}\textbf{Power}\\\textbf{Estimation}\end{tabular} & 
\begin{tabular}[c]{@{}c@{}}\textbf{Custom}\\\textbf{Topology}\end{tabular} & 
\begin{tabular}[c]{@{}c@{}}\textbf{UCIe}\\\textbf{Support}\end{tabular} & 
\begin{tabular}[c]{@{}c@{}}\textbf{2.5D/3D}\\\textbf{Mixed}\end{tabular} & 
\begin{tabular}[c]{@{}c@{}}\textbf{Hetero.}\\\textbf{Links}\end{tabular} \\
\midrule
\textbf{Booksim v2.0~\cite{jiang2013booksim}} & N$^{*}$ & Y & Y & N & N \\
\textbf{TOPAZ~\cite{abad2012topaz}}        & Y       & Limited & N & N & N \\
\textbf{Nostrum~\cite{millberg2004nostrum}}      & N       & Limited & N & N & N \\
\textbf{HeteroGarnet~\cite{bharadwaj2020kite}} & Y       & Y & Y & Y & Y \\
\bottomrule
\end{tabular}%
}
\vspace{-2mm}
\label{tab:noc_sim_selection}
\end{table}

\vspace{-2mm}
\subsection{Power Analysis and Thermal Simulation}
\label{ssec:power_and_thermal_modeling}
Thermal modeling and management of chiplet-based systems is crucial to preventing both thermal bottlenecks and potential damage to the system. Thermal simulation requires two primary inputs: a description of the physical system and a power profile over time for the chiplets. 

Figure~\ref{fig:concrete_instance} shows the flow of power and thermal data in our concrete implementation of the proposed framework. The \texttt{Global Manager} takes the workload and system configuration defined by the user as input. It then executes the simulation, following the process described in Section \ref{sec:simulation_framework}. 
%After the simulation is complete, it stores the power consumption of each compute and communication operation, \textit{along with when these operations took place and which chiplets completed the operation}. 
As the simulation progresses, it stores the power consumption of each compute and communication operation, \textit{along with when these operations took place and which chiplets completed the operation}. 
Using this information, the simulator constructs a time-based power profile for every chiplet in the system. This power profile is exported to the thermal model, shown on the right of Figure \ref{fig:concrete_instance}. 

% A wide range of thermal simulation tools have been proposed previously for both traditional monolithic chips and emerging chiplet-based systems \cite{zhang2015hotspot, sridhar20103d,yuan2021pact,pfromm2024mfit}. In this work, we select the MFIT simulator for its speed, accuracy, and adaptability to any physical system configuration \cite{pfromm2024mfit}. After receiving the power profile and taking the system configuration as input, we construct the 
% thermal profile of the system over time, as shown in Section \ref{ssec:power_thermal_analysis}.

A wide range of thermal simulation tools have been proposed previously for both traditional monolithic chips and emerging chiplet-based systems~\cite{zhang2015hotspot, sridhar20103d,yuan2021pact,pfromm2024mfit}. \rev{In this work, we use the MFIT framework~\cite{pfromm2024mfit} to analyze thermal behavior due to its accuracy, efficiency, and flexible grid resolution. MFIT supports variable spatial granularity across layers. We employ a 2×2 thermal grid per chiplet in the active layer to capture intra-chiplet temperature variations, while using coarser grids (e.g., 10×10 nodes) in passive layers such as the interposer and heat spreader. This hierarchical modeling approach enables both localized hotspot analysis and efficient simulation of system-wide heat dissipation.}
After receiving the power profile and taking the system configuration as input, we construct the 
thermal profile of the system over time, as shown in Section \ref{ssec:power_thermal_analysis}.
\vspace{-3mm}
\section{Experimental Evaluations} \label{sec:results}

\subsection{Experimental Setup}
\label{ssec:exp_setup}

\noindent\textbf{HW Configuration:}
All experiments are evaluated on a 10$\times$10 chiplet system implemented on a 2.5D silicon interposer \cite{sharma2023florets,huang2021wafer}. 
%Each chiplet is identical, making the system fully homogeneous, and each chiplet is based on the \cite{wan2022compute}.
The chiplets are configured using parameters from recent work~\cite{wan2022compute}. 
The inter-chiplet NoI uses X-Y routing. 
In Sections \ref{ssec:perf_comparison} and \ref{ssec:power_thermal_analysis}, chiplets are interconnected using a mesh NoI topology as in \cite{krishnan2021siam, shao2019simba}. 
Section \ref{ssec:additional_capabilities} presents evaluations with alternate NoI topologies and chiplet compositions to demonstrate flexibility. \rev{The chiplet composition and interconnect topology of each evaluation is shown in the left two columns of Table \ref{tab:eval_config_compare}.}

\vspace{2pt}
\noindent\textbf{Target DNN Models and Workload:}
The driver workload consists of four common DNN models: AlexNet \cite{krizhevsky2017AlexNet}, ResNet18, ResNet34, and ResNet50 \cite{he2016resnet}. These models are selected as they provide a range of memory and processing requirements, where some layers require more than one chiplet.

Each simulated workload includes 50 DNN models, uniformly sampled at random from the four DNN types described earlier. \rev{The injection rate is set 1, meaning a DNN model is loaded into the model queue on each cycle. This high injection rate maximizes the system utilization. An age-aware scheduling policy is implemented to mitigate head-of-line blocking. If a model is too large to be mapped, the arbitration policy will skip the model and attempt to map the next model instead. The arbitration policy will prioritize oldest DNN models first, however, if a model passes an age threshold, it becomes non-skipable, and will block all younger models from mapping.} Once launched, each model executes multiple back-to-back inferences before being unmapped from the system, as explained in each section. \rev{The models included in each evaluation and the number of inferences swept are shown in the right two columns of Table \ref{tab:eval_config_compare}}

\begin{table}[h]
\centering
\caption{\rev{Evaluation configuration comparison.}}
\scriptsize
\resizebox{\linewidth}{!}{%
\begin{tabular}{@{}l|c c c c@{}}
\toprule
\rev{\textbf{Evaluation}} &
\rev{\begin{tabular}[c]{@{}c@{}}\textbf{Chiplet}\\\textbf{Composition}\end{tabular}} &
\rev{\begin{tabular}[c]{@{}c@{}}\textbf{Interconn.}\\\textbf{Topol.}\end{tabular}} &
\rev{\textbf{Models}} &
\rev{\begin{tabular}[c]{@{}c@{}}\textbf{\# Inf.}\\\textbf{Swept}\end{tabular}} \\
\midrule
\rev{\textbf{\begin{tabular}[c]{@{}l@{}}Non-pipelined\\Comparison\end{tabular}}} &
\rev{Homog.} \cite{wan2022compute} & \rev{Mesh} \cite{krishnan2021siam, shao2019simba} &
\rev{\begin{tabular}[c]{@{}c@{}}AlexNet\\ResNets\end{tabular}} &
\rev{\begin{tabular}[c]{@{}c@{}}10\\~\end{tabular}} \\
\cmidrule(lr){1-5}
\rev{\textbf{\begin{tabular}[c]{@{}l@{}}Pipelined\\Comparison\end{tabular}}} &
\rev{Homog.} \cite{wan2022compute} & \rev{Mesh} \cite{krishnan2021siam, shao2019simba} &
\rev{\begin{tabular}[c]{@{}c@{}}AlexNet\\ResNets\end{tabular}} &
\rev{\begin{tabular}[c]{@{}c@{}}1, 3, 5\\10, 20\end{tabular}} \\
\cmidrule(lr){1-5}
\rev{\textbf{\begin{tabular}[c]{@{}l@{}}Heterog.\\Comparison\end{tabular}}} &
\rev{Heterog.} \cite{wan2022compute, andrulis2023raella} & \rev{Mesh} \cite{krishnan2021siam, shao2019simba} &
\rev{\begin{tabular}[c]{@{}c@{}}AlexNet\\ResNets\end{tabular}} &
\rev{\begin{tabular}[c]{@{}c@{}}1, 3, 5\\10, 20\end{tabular}} \\
\cmidrule(lr){1-5}
\rev{\textbf{\begin{tabular}[c]{@{}l@{}}Floret \cite{sharma2023florets}\\Comparison\end{tabular}}} &
\rev{Homog.} \cite{wan2022compute} & \rev{Floret} \cite{sharma2023florets} &
\rev{\begin{tabular}[c]{@{}c@{}}AlexNet\\ResNets\end{tabular}} &
\rev{\begin{tabular}[c]{@{}c@{}}1, 3, 5\\10, 20\end{tabular}} \\
\cmidrule(lr){1-5}
\rev{\textbf{\begin{tabular}[c]{@{}l@{}}Transformer\\Evaluation\end{tabular}}} &
\rev{Homog.} \cite{wan2022compute} & \rev{Mesh} \cite{krishnan2021siam, shao2019simba} &
\rev{\begin{tabular}[c]{@{}c@{}}ViT-B\\~\end{tabular}} &
\rev{\begin{tabular}[c]{@{}c@{}}1, 2, 5\\10, 20\end{tabular}} \\
\bottomrule
\end{tabular}%
}
\vspace{-2mm}
\label{tab:eval_config_compare}
\end{table}

\vspace{2pt}
\noindent\textbf{Simulation Parameters:}
\rev{To balance simulation fidelity and runtime, \frameworkName{} uses a configurable time step of 1$\mu$s during co-simulation. This granularity was selected based on typical execution durations of DNN layer compute and communication phases, which range from several microseconds to hundreds of microseconds in our setup. We validated that a smaller time step does not further improve the simulation results.} \rev{A warm-up and cool-down time are applied to each simulation, during which simulation statistics are not collected. Both warm-up and cool-down times are set to 1 ms.}

\vspace{2pt}
\noindent\textbf{Model to System Mapping:}
To map DNN layers to the chiplet-based system, we use a nearest-neighbor mapping strategy inspired by Simba \cite{shao2019simba}. This mapper ensures that consecutive layers are mapped to spatially close chiplets, minimizing communication overhead. 

\vspace{2pt}
\noindent\textbf{Baseline Comparisons:}
We consider two baseline approaches for comparison to the \frameworkName{} simulation framework. First, we consider the network-only approach utilized in NoI exploration works \cite{sharma2023florets,iff2023hexamesh}. In this baseline, only the network is simulated, and the compute simulation is omitted. We refer to this baseline as "\textit{Comm. Only}". Second, we consider the approach implemented in SIAM, HISIM, and other more comprehensive works \cite{krishnan2021siam, wang2025hisim}. We refer to this baseline as "\textit{Comm. + Compute}". In this baseline approach, compute and communication are decoupled and simulated for each layer of each DNN model individually. In both baselines, only a single model is present in the system at a time during simulation. 
The same nearest-neighbor mapper that is utilized in \frameworkName{} is used to map models for simulation using the baseline methods.

\vspace{-3mm}
\subsection{Performance Comparisons vs Baseline Approaches} \label{ssec:perf_comparison}

This section presents results using a mesh NoI and homogeneous chiplets with non-pipelined and pipelined operation.

\subsubsection{\textbf{Non-pipelined Operation}} \label{ssec:baseline_vs_cosim}
In this mode, \textit{there may be multiple DNN model instances that cause contention in the NoI, but layers of a given DNN model are executed one at a time without pipelining}. 
This design choice does not fully exploit the system and leads to low utilization, but it can be preferred to simplify the hardware and software designs.
Once launched, each model performs 10 back-to-back inferences before leaving the system.

\begin{table}[h]
\centering
\vspace{-1mm}
\caption{Impact of co-simulation relative to baseline approaches for non-pipelined operation.}
\vspace{-1mm}
\label{tab:non-pipelined_comparison}
\begin{tabular}{c|cc}
\toprule
\textbf{DNN Model} & 
\multicolumn{2}{c}{\textbf{Percent Inaccuracy}} \\
\cmidrule(lr){2-3}
 & \textbf{\textit{Comm. Only}} & \textbf{\textit{Comm. + Compute}} \\
\midrule
\textbf{ResNet18}  & 74\%  & 8\%  \\
\textbf{ResNet34}  & 70\%  & 11\% \\
\textbf{ResNet50}  & 22\%  & 9\%  \\
\textbf{AlexNet}   & 33\%  & 24\% \\
\bottomrule
\end{tabular}
\vspace{-1mm}
\end{table}

Table~\ref{tab:non-pipelined_comparison} compares the accuracy of the two baseline methods in predicting end-to-end inference latency against \frameworkName{}.
%, where the percentage inaccuracy is relative to the results produced by \frameworkName{}.
The \textit{Comm. Only} 
baseline underestimates the latency by 22\% (for ResNet50) to 74\% (for ResNet18). These errors result from (1) the lack of compute simulation and (2) network traffic contention modeling in communication simulation.

The modeling error decreases by accounting for communication in the \textit{Comm.~+~Compute} baseline, as expected. 
The largest error occurs for AlexNet at 24\%, while the smallest error occurs for ResNet18 at 8\%.  These results show that for non-pipelined operation, disaggregated \textit{Comm.~+~Compute} simulation is a reasonable approximation of when the DNN model layers are non-pipelined. However, this inaccuracy increases dramatically with pipelining, as discussed next. 

\subsubsection{\textbf{Pipelined Operation}}
A more optimized implementation pipelines the layers of each DNN model. Once a chiplet completes processing a layer and sends out the activations, it immediately starts the next inference round with the new input, while another chiplet processes its output. Hence, multiple layers of a given DNN model run in parallel.

% Figure comparing the system utilization to the average execution latency
\begin{figure}[t]
    \centering
    \includegraphics[width=1\linewidth]{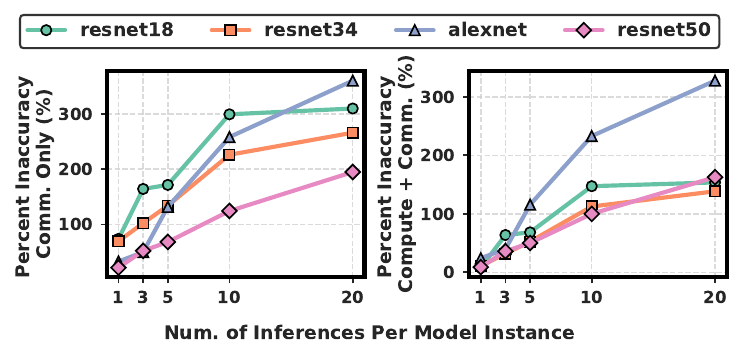}
    \caption{The baseline approaches significantly underestimate the end-to-end inference time as the number of inferences per model (hence utilization) increases.}
    \label{fig:latency_vs_num-inputs}
    \vspace{-2mm}
\end{figure}

% Figure comparing the baseline to co-simulation results pipelined with 10 inputs
\begin{figure}[t]
    \centering
    \includegraphics[width=1\linewidth]{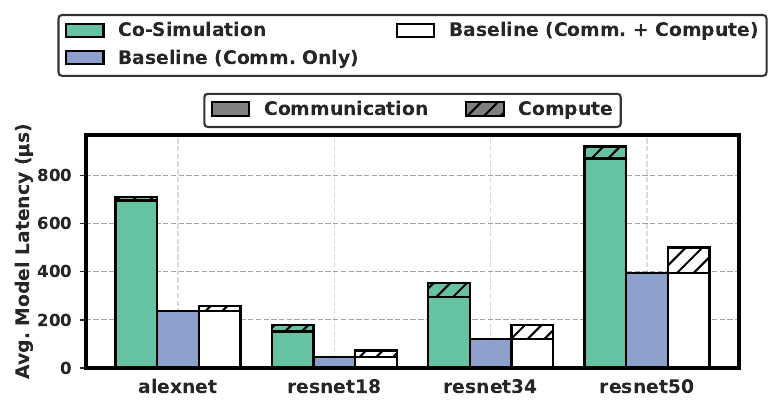}
    \caption{Pipelined performance comparisons with 10 inferences per model.}
    \label{fig:10_input_comparison}
    \vspace{-2mm}
\end{figure}

Figure~\ref{fig:latency_vs_num-inputs} compares the end-to-end inference latencies reported by \frameworkName{} and baseline approaches as a function of inferences per DNN model. When the DNN models leave the system after only one inference, they do not exploit pipelining and operate like the non-pipelining case. 
The number of inferences repeated per model instance increases the number of parallel inferences, i.e., pipelining, and utilization.
Hence, this experiment enables analyzing the accuracy benefits of \frameworkName{} as a function of system utilization.

Aligned with the non-pipelined results, the \textit{Comm. Only} baseline has significant inaccuracy, even with one or few inferences per model, since it fails to account for computation and network contention.
The inaccuracy increases with the number of inferences per model instance. 
The error starts to saturate at 20 inferences, where AlexNet exceeds 340\% latency inaccuracy and other models show 200-300\% error, as the system reaches maximum utilization. 
Even at one inference per model instance, where pipelining is effectively disabled, ResNet18 and 34 show 74\% and 69\% error, respectively, while ResNet50 and AlexNet show lower error at 22\% and 33\% error, respectively.

The \textit{Comm.~+~Compute} baseline exhibits a similar trend, though with lower underestimation. 
Again, the error starts saturating at the final data point of 20 inferences/model instance as more parallelism no longer increases the utilization. 
AlexNet experiences up to 320\% modeling error, while the errors of the ResNet models range from 100\% to 200\%. 
There is a smaller (less than 30\%) difference between using the baseline approach or co-simulation with a single inference for each model (effectively no pipelining), aligned with previous results.

Figure~\ref{fig:10_input_comparison} plots the average compute and communication times for each model to analyze the root cause of the underestimation. 
Since multiple layers of each DNN model run in parallel with pipelining, large inter-chiplet communication volume and NoI utilizations lead to higher traffic contention and communication time.
In these simulations, the communication time dominates the total inference time for this particular hardware configuration, since the selected chiplets have fast processing speeds. 
\textit{Comm. Only} or \textit{Comm.~+~Compute} baselines are inaccurate even under this condition since the proposed co-simulation approach is required to capture the network traffic timing and contention. 
The next section adds slower chiplets to the system to analyze scenarios where computation takes up a larger portion of the total processing time.

% Non-mesh NoI and Heterogeneous results. 
\vspace{-3mm}
\subsection{Demonstrations with Different NoIs and Chiplet Types} \label{ssec:additional_capabilities}

% Section on Heterogeneous System Experiment
\subsubsection{\textbf{Heterogeneous Chiplets}} \label{sssec:heterogeneous_chiplets}
%In addition to alternate topologies, 
\frameworkName{} supports systems composed of heterogeneous chiplets. To demonstrate this ability, we modify the original homogeneous 100 chiplet system to create a system with 50 chiplets of the original type and 50 of a different type based on a separate IMC work \cite{andrulis2023raella}, with pipelining enabled. The chiplets are laid out in an alternating pattern, such that each chiplet is neighbored by chiplets of the other type. 

The computation time increases significantly, reaching 42\% of the total execution time for Resnet18, 44\% for Resnet34, and 54\% for Resnet50. 

% \rev{Two different heterogeneous system configurations were considered for evaluation. First, a clustered system, where the system is divided in half and each half is made up of a different chiplet type. Second, an alternating system, where the type of each chiplet alternates, forming an alternating pattern.}

% \rev{A key issue arises with the clustered system. When performing baseline evaluations, the model is mapped into a completely empty system, meaning the model could be mapped to any chiplet. However, the nearest-neighbor mapper implemented for our evaluations has two objectives: utilize the highest performance chiplets, and map layers spatially close. Additionally, the considered DNN models typically require on the order of 5-40 chiplets to map. Therefore, in the clustered system, with all chiplets available and nearest-neighbor mapping applied, these baseline models will be mapped only to the highest performing chiplet cluster and will not require any of the lower performance chiplets. With this mapping, the baseline results produced by the heterogeneous system will be identical to the homogeneous system which is made up of \textbf{only} higher performance chiplets.}

% \rev{To mitigate this issue, we instead perform evaluations on the alternating heterogeneous system. In this system, the mapper is forced to choose between keeping layers spatially close and mapping to high performance chiplets, providing a better representation of the actual system performance.}

Table \ref{tab:heterogeneous_comparison} compares the percent inaccuracy in results generated using the \textit{Comm.~+~Compute} baseline as compared to \frameworkName{} results. 
We exclude the \textit{Comm. Only} baseline due to its previously demonstrated inaccuracy. 
The baseline method exhibits substantial inaccuracy in simulating this heterogeneous system. Again, the minimum error occurs with one inference per model instance, with a minimum error of 9\% for AlexNet and a maximum error of 146\% for ResNet18. 
As before, error increases substantially with the number of inferences. Maximum error occurs at the maximum number of inferences per model instance, with ResNet18 showing the largest inaccuracy of the baseline method at 632\%, while ResNet50 shows the smallest at 233\%. 

\begin{table}[h]
\centering
\caption{Percent inaccuracy relative to baseline latency across inference counts using a heterogeneous system.}
\begin{tabular}{c|cccc}
\toprule
\textbf{Num. of Inferences} & 
\multicolumn{4}{c}{\textbf{Percent Inaccuracy}} \\
\cmidrule(lr){2-5}
 & \textbf{ResNet18} & \textbf{ResNet34} & \textbf{ResNet50} & \textbf{AlexNet} \\
\midrule
1  & 146\% & 52\%  & 49\%  & 9\%   \\
3  & 310\% & 113\% & 69\%  & 39\%  \\
5  & 353\% & 211\% & 81\%  & 107\% \\
10 & 407\% & 250\% & 155\% & 224\% \\
20 & 632\% & 423\% & 233\% & 337\% \\
\bottomrule
\end{tabular}
\label{tab:heterogeneous_comparison}
\end{table}

% % Figure comparing the baseline to co-simulation results pipelined with 10 inputs for heterogeneous system
% \begin{figure}[h]
%     \centering
%     \includegraphics[width=1\linewidth]{figures/simulation_approach_comparison_results/heterogeneous/approach_stacked_runtime_comparison.pdf}
%     \caption{\todo{Same plot as fig 7, but for heterogeneous system.}}
%     \label{fig:10_input_comparison_heterogeneous}
%     \vspace{-2mm}
% \end{figure}

% Section on Floret Experiment
\subsubsection{\textbf{Alternate Topologies}} \label{sssec:alternate_topologies}

Our co-simulation framework supports arbitrary NoI topologies beyond the mesh topology used in the previous section. We demonstrate this capability using the recently proposed Floret topology on the same 100-chiplet system~\cite{sharma2023florets}, which is designed to optimize DNN execution by aligning the NoI structure with DNN data flows.
Workloads and system configurations remain identical, and pipelining is enabled.  

Table~\ref{tab:topology_comparison} summarizes the percentage error of the \textit{Comm.~+~Compute} baseline only relative to \frameworkName{} since the \textit{Comm. Only} baseline is shown to be inferior. 
As observed with the mesh topology, the baseline error grows with increasing inferences per model. 
At one inference per model, where the pipelining effects are negligible, the maximum error is 17\% (AlexNet). 
The error increases with more inference per model (hence, utilization). At 20 inferences per model, errors escalate dramatically, reaching 120\% for ResNet34 and 337\% for AlexNet.

\begin{table}[t]
\centering
\caption{Percent inaccuracy relative to baseline latency across inference counts using the Floret NoI.}
\begin{tabular}{c|cccc}
\toprule
\textbf{Num. of Inferences} & 
\multicolumn{4}{c}{\textbf{Percent Inaccuracy}} \\
\cmidrule(lr){2-5}
 & \textbf{ResNet18} & \textbf{ResNet34} & \textbf{ResNet50} & \textbf{AlexNet} \\
\midrule
1   & 14\%  & 13\%  & 9\%   & 17\%   \\
3   & 56\%  & 42\%  & 30\%  & 24\%   \\
5   & 68\%  & 53\%  & 49\%  & 131\%  \\
10  & 94\%  & 102\% & 103\% & 235\%  \\
20  & 126\% & 120\% & 128\% & 337\%  \\
\bottomrule
\end{tabular}
\vspace{-2mm}
\label{tab:topology_comparison}
\end{table}

% Section on Power and Thermal information provided by the tool
\subsection{Power and Thermal Analysis} \label{ssec:power_thermal_analysis}
\begin{figure}[b]
    \vspace{-5mm}
    \centering
    \includegraphics[width=\linewidth]{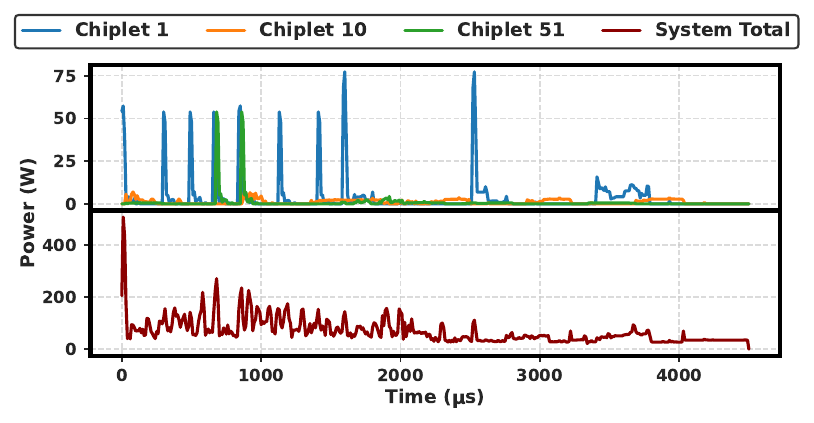}
    \caption{Upper plot shows per-chiplet power over time of 3 selected chiplets. Lower plot shows System Total power over time.}
    \label{fig:power_over_time}
\end{figure}

In addition to performance results, each simulation run with our co-simulation framework generates power consumption profiles over time for every chiplet in the system. 
Figure~\ref{fig:power_over_time} presents the power profiles of three representative chiplets executing a full workload (upper plot) along with the total system power consumption (lower plot). The per-chiplet profiles clearly show transient power spikes, such as those observed in chiplets 1 and 51. Our implementation captures power at 1$\mu$s granularity, enabling accurate modeling of short-lived power fluctuations that would be missed with coarser methods.

% Plot of per-chiplet power over time

The lower plot of Figure \ref{fig:power_over_time} shows the total system power. Smaller models, which require fewer chiplet resources, are mapped and execute earlier in the simulation as they are less likely to be blocked from mapping, while larger models, which require more memory, execute later in the simulation as more memory becomes free. This behavior is accurately reflected in the power consumption profile, which shows a significant power spike early in the simulation. It is followed by rapidly fluctuating power levels before power stabilizes at around the 50~W range. 

\begin{figure}[t]
    \centering
    \includegraphics[width=0.3\textwidth]{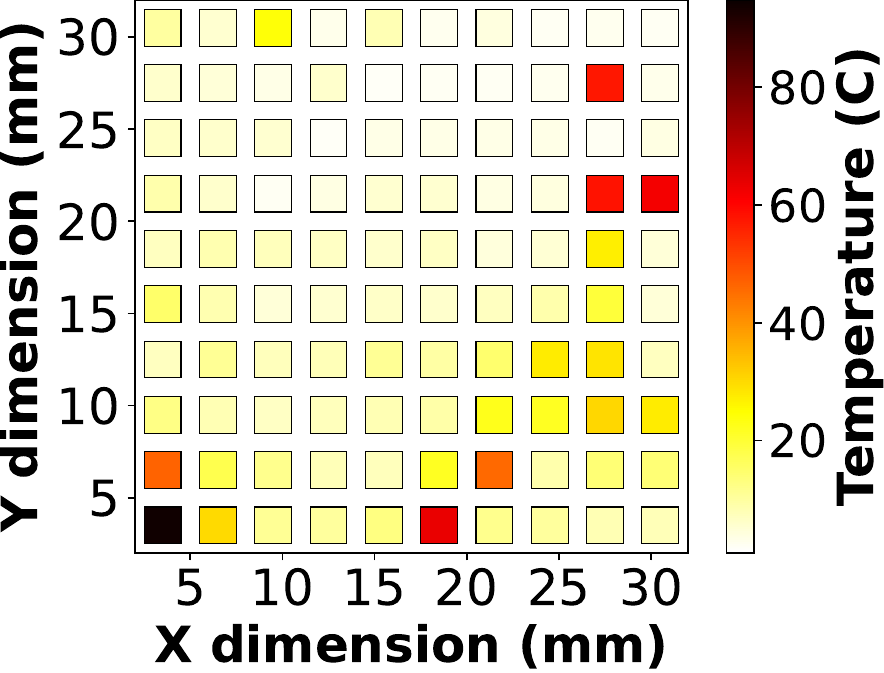}
    \caption{Heatmap at the end of simulation.}
    \label{fig:heatmap}
    \vspace{-2mm}
\end{figure}

These detailed power traces are then used as inputs to our integrated fast thermal model, as described in Section \ref{ssec:power_and_thermal_modeling}. 
The thermal model analyzes the transient temperature and steady-state thermal profile of each chiplet using this input. 
Illustrative thermal maps are shown in Figure \ref{fig:heatmap}, where darker colors show which chiplets reach higher temperatures. This heatmap allows the user to visualize hotspots within the chiplet-based system. 
The power-thermal analysis capability allows precise analysis of how workload execution affects chiplet temperatures, providing critical insights for thermal management and reliability assessment.

\subsection{\rev{Transformer Model Demonstration}}
\label{ssec:transformer_model}

\rev{The} \frameworkName{} \rev{framework is capable of simulating transformer models in addition to the previously demonstrated CNN models. To demonstrate this ability, we simulate the ViT-B/16} \cite{vision_transformer} \rev{model on the same system used to simulate CNN models in Section 5.b. This system has 100 chiplets of identical type with a mesh interconnect with the following modifications:}
\begin{itemize}
    \item \rev{The four corner chiplets are designated as I/O chiplets, responsible for hosting and distributing model weights.}
    \item \rev{The system adopts weight-stationary in-memory compute (IMC): weights are loaded once at the beginning and reused across inferences.}
    \item \rev{We simulate single-model execution with input pipelining, without concurrent model instances, due to the memory footprint of ViT-B/16.}
\end{itemize}

\begin{figure}[h]
    \centering
    \includegraphics[width=0.5\textwidth]{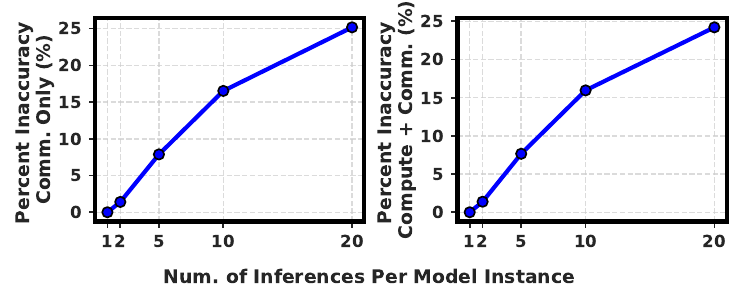}
    \caption{\rev{Baseline vs} \frameworkName{} \rev{results for ViT-B execution.}}
    \label{fig:vit_comparison}
    \vspace{-2mm}
\end{figure}

\rev{Due to the weight storage requirements of a single Vit-B model, a single instance of the model is simulated without the parallel execution of any additional models. As with previous shown results, pipelining of inputs is enabled in this example. 
Figure} \ref{fig:vit_comparison} \rev{shows the impact of increasing the number of inferences completed by the ViT-B model. The left subplot shows the difference between the results of} \frameworkName{} \rev{and the communication only baseline, while the right shows the difference between} \frameworkName{} \rev{and the compute + communication baseline. There is no difference at one inference per model since only a single model is running. The difference increases and reaches 25\% at 20 inferences per model for the communication only baseline, and 24\% for the compute + communication baseline. The smaller difference in results between the baseline and} \frameworkName{} \rev{results for this experiment is driven by two factors. First, for a single inference, weight loading time is dominant, taking about three times longer than the execution of the model, and is accounted for in both the baseline and} \frameworkName{} \rev{results. The impact of this fixed overhead becomes relatively less as the number of inferences increases. At larger numbers of inferences, the difference between baseline and} \frameworkName{} \rev{results is driven by contention between pipelined inputs and parallel model execution. As no parallel models exist in this simulation, only pipelined input contention causes inaccuracy, explaining the relatively lower difference between} \frameworkName{} \rev{and baseline results. The relatively smaller difference between the communication only and compute + communication baseline is due to the dominance of communication latency during ViT-B execution, leading to a very small difference between these two baselines.} 

\subsection{\rev{Hardware Validation}}\label{ssec:hw_validation}

\noindent\textbf{\rev{Hardware Setup:}}
\rev{We further validate the proposed CHIPSIM framework against hardware measurements collected on an AMD Ryzen Threadripper PRO 7985WX (64 cores) platform~\cite{amd_trpro_7985wx_datasheet}. This platform contains 8 CCD (compute complex die) chiplets and 1 IOD (I/O die) responsible for chiplet-to-chiplet and DRAM communication. Each CCD contains 8 cores and connects to the IOD via Global Memory Interconnect (GMI3) links, while the IOD interfaces with DDR5 memory. Notably, the GMI3 fabric provides asymmetric bandwidth per direction (32 B/cycle for read, 16 B/cycle for write at 1.733 GHz), corresponding to $\sim$55 GB/s read and $\sim$27.7 GB/s write per link.} 

\vspace{2pt}
\noindent\textbf{\rev{DL Workload Modeling on Chiplet Platforms:}}
\rev{DL workloads consist of interleaving computation within a layer/module (e.g., fully connected layer, attention block) and communication of the produced activation outputs. Computations are confined within a processing unit (like a chiplet), while communication is between chiplets and memory. To mimic execution of DL workloads on IMC platforms, we develope a macro-kernel workload that integrates load, compute, and store phases in a configurable loop. Our macro-kernels allow explicit compute and communication scheduling with parameters that control their duration. Each CCD in the AMD Ryzen Threadripper PRO 7985WX platform executes this workload independently, emulating multiple DNNs executing concurrently on different chiplets, similar to our IMC chiplet setup.}

\rev{To simulate this system in CHIPSIM, we first implement the same topology present in the hardware system in HeteroGarnet by configuring heterogeneous links that match the measured read/write bandwidth of each CCD as well as the IOD-to-DDR interface. This setup allows us to capture inter-chiplet and memory communication. CiMLoop is replaced with an analytical compute model created from the FLOPs/sec measured for each chiplet while computing micro-kernel benchmarks. This approach ensures that the compute throughput in CHIPSIM faithfully reflects the actual platform characteristics. CHIPSIM provides a realistic co-simulation of parallel DL workloads on a chiplet-based CPU by integrating both the bandwidth-calibrated communication model and the analytically derived compute model.}

\begin{figure}[t]
    \centering
    \includegraphics[width=1\linewidth]{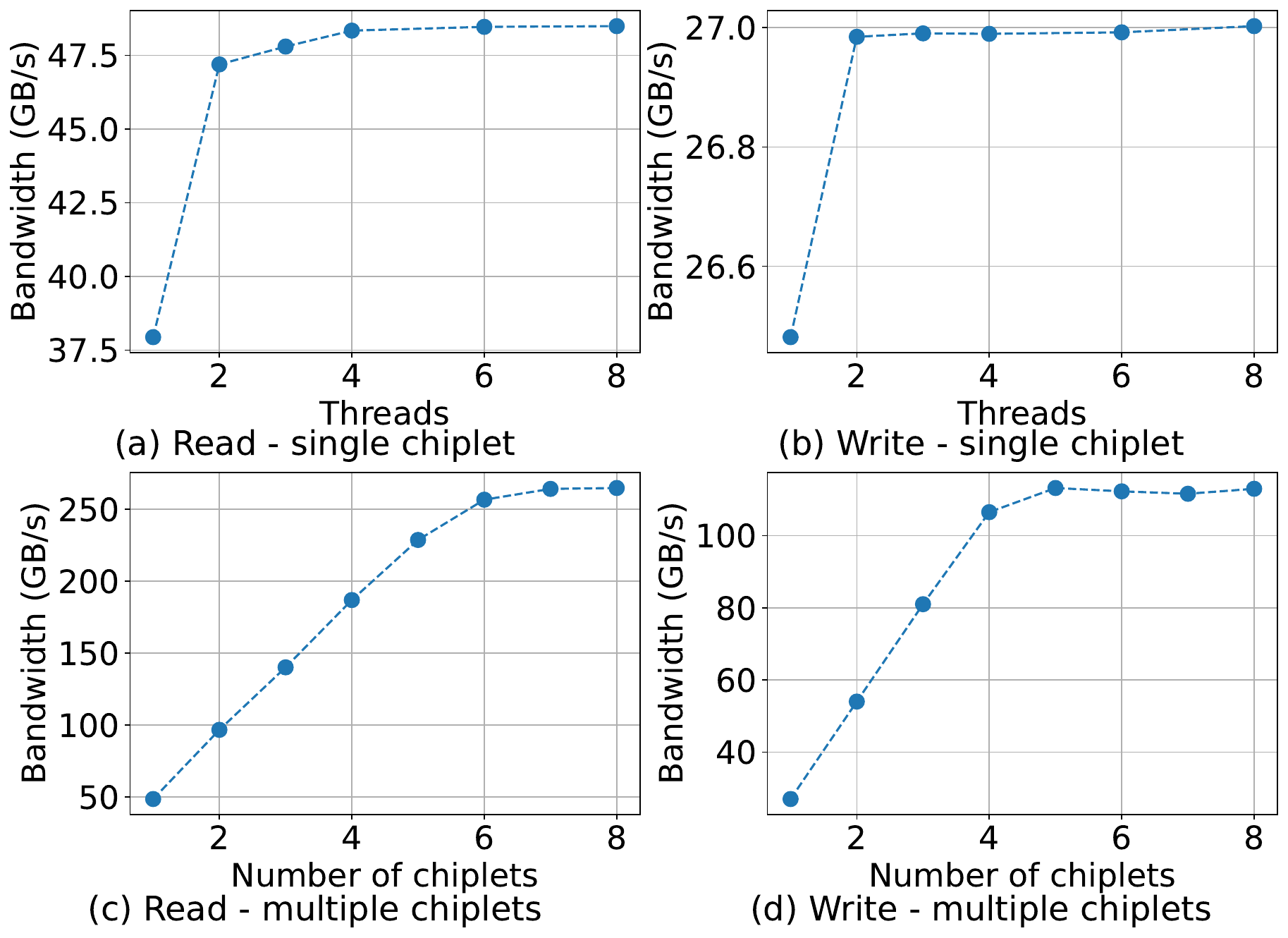}
    \caption{\rev{Measured bandwidth results on the AMD Ryzen Threadripper PRO 7985WX platform. (a) Single-CCD read bandwidth as function of active threads. (b) Single-CCD write bandwidth as a function of active threads. (c) Aggregate read bandwidth as a function of active CCDs, each running 8 threads. (d) Aggregate write bandwidth as a function of active CCDs, each running 8 threads.}}
    \label{fig:amd_bw_plot}
    \vspace{-2mm}
\end{figure}

\vspace{2pt}
\noindent\textbf{\rev{Hardware Profiling:}}
\rev{For ground-truth measurements, we first employ the LIKWID microkernel suite~\cite{treibig2010likwid}. Load–store benchmarks exclude computation to isolate communication costs. Figures~\ref{fig:amd_bw_plot}(a)(b) show single-CCD read and write bandwidth scaling with threads, saturating at $\sim$49 GB/s and $\sim$27 GB/s, respectively. These values achieve $\sim$90\% (read) and $\sim$98\% (write) of the theoretical GMI3 peak. Figures~\ref{fig:amd_bw_plot}(c)(d) report aggregate bandwidth across multiple CCDs, with all 8 threads per CCD enabled. Read bandwidth scales to $\sim$270 GB/s, while write bandwidth saturates around ~115 GB/s, demonstrating DDR congestion as more CCDs become active. These measurements approach the advertised DDR5 peak bandwidth of $\sim$330 GB/s, reaching $\sim$83\% utilization. \emph{This close alignment confirms that our profiling captures the characteristics of the underlying platform and provides a reliable ground truth for evaluating ChipSim.}}

\vspace{2pt}
\noindent\textbf{\rev{Validation of Chipsim Results:}}
\rev{
Three identical CNN workloads are executed on both the AMD Ryzen Threadripper PRO 7985WX platform and our simulator. For the first scenario, a single chiplet executes a single instance of AlexNet. The second scenario executes two instances of AlexNet on two different chiplets. Finally, the third scenario executes one instance of AlexNet, ResNet18, ResNet34, and ResNet50 each on four different chiplets. %For each model instance, a single forward inference pass is executed.
Table~\ref{tab:chipsim_vs_hw} reports the percentage difference between measured and simulated times for each model. The third column shows the error in execution latency for each individual model, while the last column shows the average error across all models. 
\frameworkName{} results align closely with hardware measurements, showing a maximum average error of only 2.51\% for the four chiplet scenario and a minimum error of 0.75\% for the two chiplet scenario. Per-model maximum error is less than 6\%, seen when comparing results of simulating ResNet50.
% For instance, running a single instance of AlexNet on one chiplet yields only a 0.77\% deviation, while executing two parallel instances across separate chiplets results in differences of 0.64\% and 0.86\%, respectively. Expanding the workload to include four different types of DNN model results in the same minimum error of 0.65\% for AlexNet and a maximum error of 8.66\% error for ResNet50.
These results demonstrate that CHIPSIM not only enables design space exploration for new architectures but also provides highly accurate predictions against existing hardware platforms.
}

% HW validation results table
\begin{table}[h]
\centering
\caption{\rev{Percent difference of CHIPSIM results as compared to hardware results.}}
\renewcommand{\arraystretch}{1.2}
\begin{tabular}{l|l|c|c}
\toprule
\textbf{\rev{Num. of Chiplets}} & \textbf{\rev{Models}} & \textbf{\rev{\% Diff. from HW}} & \textbf{\rev{Avg. \% Diff.}} \\
\midrule
\multirow{1}{*}{\rev{One Chiplet}} & \rev{AlexNet} & \rev{0.77\%} & \rev{0.77\%} \\
\hline
\multirow{2}{*}{\rev{Two Chiplets}} & \rev{AlexNet (1)} & \rev{0.64\%} & \multirow{2}{*}{\rev{0.75\%}} \\
                                    & \rev{AlexNet (2)} & \rev{0.86\%} &  \\
\hline
\multirow{4}{*}{\rev{Four Chiplets}} & \rev{AlexNet}  & \rev{1.29\%} & \multirow{4}{*}{\rev{2.51\%}} \\
                                     & \rev{ResNet18} & \rev{0.84\%} &  \\
                                     & \rev{ResNet34} & \rev{2.19\%} &  \\
                                     & \rev{ResNet50} & \rev{5.74\%} &  \\
\bottomrule
\end{tabular}
\label{tab:chipsim_vs_hw}
\end{table}

% \vspace{-2mm}
\subsection{Execution Time Analysis}

Table \ref{tab:simulation_runtime} lists the average runtime per DNN model for each simulation approach. The \frameworkName{} framework has an execution time of 12.6 minutes per model. In comparison, the \textit{Comm.~+~Compute} baseline takes 12.2 minutes per model. The small increase in execution time shown by \frameworkName{} is due to the additional overhead required for accurately tracking the state of the system, making mapping decisions, and coordinating the execution of component simulators. 
These results show that substantial accuracy improvement of \frameworkName{} comes at a moderate increase in the execution time.

% Results show that the cycle-accurate simulation employed by gem5 is completely infeasible for any sort of rapid simulation due to its massive runtime, taking weeks to fully simulate a single DNN model. Baseline simulation and \frameworkName{} show very similar execution times, with \frameworkName{} requiring slightly longer to simulate a single model. This difference is due to the additional overhead required for accurately tracking the state of the system, making mapping decisions, and coordinating the execution of component simulators. 

Our \frameworkName{} implementation already performs cycle-accurate communication simulation using Hetero Garnet~\cite{bharadwaj2020kite}. 
Further accuracy improvement can be achieved by making computation simulations also cycle-accurate. Since the most common choice, gem5~\cite{lowe2020gem5}, cannot simulate the system considered in this work due to the use of IMC, we use another prior work that simulates DNN workloads on non-IMC systems using gem5~\cite{ramadas2024simulation}. 
This prior work demonstrates that performing cycle-accurate simulation of DNN models in gem5 takes on the order of weeks.
This massive execution time makes cycle-accurate computation-communication simulations infeasible for rapid evaluations, 
while \frameworkName{} achieves speed, accuracy, and modularity through carefully optimized co-simulations.

\begin{table}[h]
\centering
\caption{Simulation runtime comparison across methods.}
\begin{tabular}{l|c}
\toprule
\textbf{Simulation Method} & \textbf{Avg. Execution Time per Model} \\
\midrule
\textbf{\frameworkName{} (This Work)} & 12.6 min \\
\textbf{Comm.~+~Compute Baseline}     & 12.2 min \\
\textbf{Cycle-Accurate (gem5)}        & Weeks \cite{ramadas2024simulation} \\
\bottomrule
\end{tabular}
\vspace{-2mm}
\label{tab:simulation_runtime}
\end{table}

\section{Conclusion} \label{sec:conclusion}

In this work, we introduce \frameworkName{}, a fast and accurate co-simulation framework for chiplet-based systems executing deep neural networks. By modeling computation and communication together with a coherent global timeline, \frameworkName{} captures key effects like pipelining and network contention that are missed by traditional methods. Evaluations show that existing approaches can wrongly estimate latency by over 800\% under realistic workloads. In contrast, \frameworkName{} delivers high accuracy with fast runtime and supports detailed power and thermal analysis, making it a valuable tool for designing and optimizing chiplet-based systems.

\bibliographystyle{IEEEtran}
\bibliography{references/refs}

% ====== Author Biographies (IEEEtran) ======

% Example with photo
\begin{IEEEbiography}[{\includegraphics[width=1in,height=1.25in,clip,keepaspectratio]{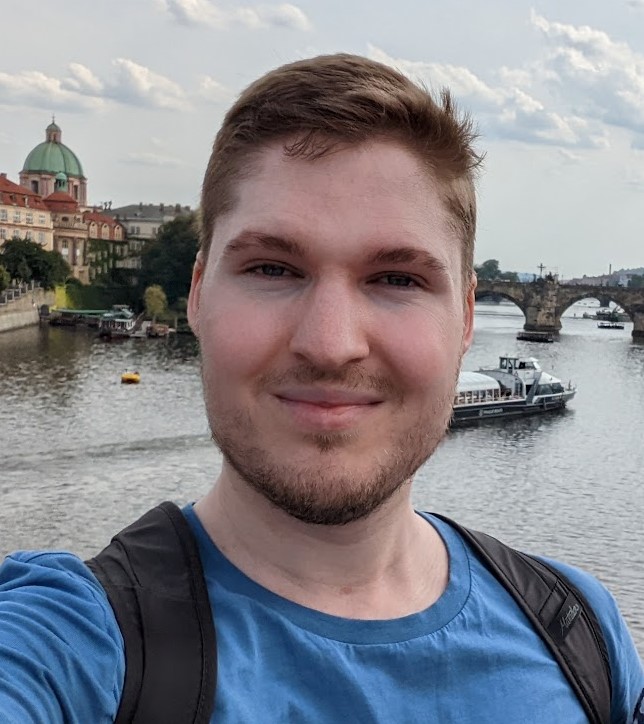}}]{Lukas Pfromm}
(Student Member, IEEE) is currently pursuing the Ph.D. degree in Computer Engineering at University of Wisconsin Madison, Madison, WI, USA. His research interests include chiplet system performance modeling, thermal modeling, and thermal management. 
% \vspace{0.25em}\noindent\emph{ORCID:} 0000-0000-0000-0000 \quad \emph{Email:} pfromm@wisc.edu
\end{IEEEbiography}

% Example with photo
\begin{IEEEbiography}[{\includegraphics[width=1in,height=1.25in,clip,keepaspectratio]{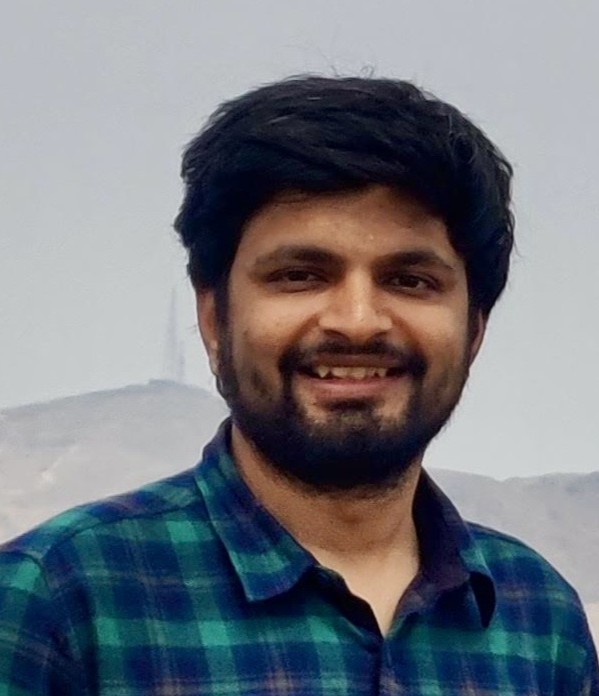}}] {Alish Kanani} (Student Member, IEEE) is currently pursuing the Ph.D. degree in Computer Engineering at University of Wisconsin Madison, Madison, WI, USA. His research interests include runtime thermal management and co-optimized system design. 
\end{IEEEbiography}

% Example with photo
\begin{IEEEbiography}[{\includegraphics[width=1in,height=1.25in,clip,keepaspectratio]{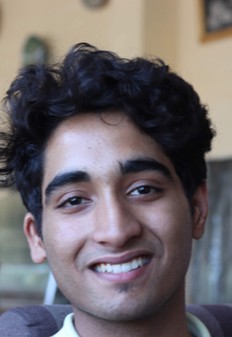}}]{Harsh Sharma}
(Graduate Student Member, IEEE) is currently pursuing the Ph.D. degree in Computer Engineering at Washington State University, Pullman, WA, USA. His research interests include heterogenous integration, and ML for computer-aided design.
% \vspace{0.25em}\noindent\emph{Email:} harsh.sharma@wsu.edu
\end{IEEEbiography}

% Example with photo
\begin{IEEEbiography}[{\includegraphics[width=1in,height=1.25in,clip,keepaspectratio]{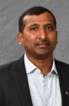}}]{Janardhan Rao Doppa} (Senior Member, IEEE) received the Ph.D. degree in computer science from Oregon State University, Corvallis, OR, USA, in 2014. He is the Huie-Rogers Endowed Chair Associate Professor in Computer Science at Washington State University, Pullman, WA, USA.
% \vspace{0.25em}\noindent\emph{Email:} jana.doppa@wsu.edu
\end{IEEEbiography}

% Example with photo
\begin{IEEEbiography}[{\includegraphics[width=1in,height=1.25in,clip,keepaspectratio]{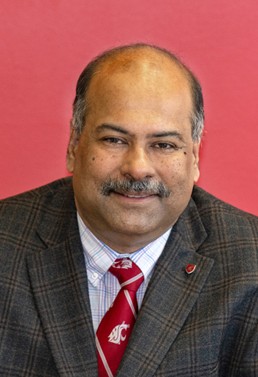}}]{Partha Pratim Pande}
(Fellow, IEEE) received the Ph.D. degree in electrical and computer engineering from the University of British Columbia, Vancouver, BC, Canada, in 2005. He is a professor and a holder of the Boeing Centennial Chair of Computer Engineering with the School of EECS, Washington State University, Pullman, WA, USA.
% \vspace{0.25em}\noindent\emph{Email:} pande@wsu.edu
\end{IEEEbiography}

% Example with photo
\begin{IEEEbiography}[{\includegraphics[width=1in,height=1.25in,clip,keepaspectratio]{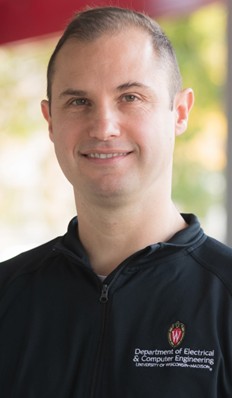}}]{Umit Y. Ogras}
(Fellow, IEEE) received the Ph.D. degree from Carnegie Mellon University, Pittsburgh, PA, USA, in 2007. He is currently an Associate Professor with the Department of ECE at University of Wisconsin–Madison, Madison, WI, USA.
% \vspace{0.25em}\noindent\emph{Email:} uogras@wisc.edu
\end{IEEEbiography}

\end{document}